\documentclass[sn-apa]{sn-jnl}% APA Reference Style
\usepackage{graphicx}%
\usepackage{multirow}%
\usepackage{amsmath,amssymb,amsfonts}%
\usepackage{amsthm}%
\usepackage{mathrsfs}%
\usepackage[title]{appendix}%
\usepackage{xcolor}%
\usepackage{textcomp}%
\usepackage{manyfoot}%
\usepackage{booktabs}%
\usepackage{algorithm}%
\usepackage{algorithmicx}%
\usepackage{algpseudocode}%
\usepackage{listings}%
\usepackage[T1]{fontenc}
\usepackage[utf8]{inputenc}
\usepackage{xurl}
\theoremstyle{thmstyleone}%
\theoremstyle{thmstyletwo}%

\theoremstyle{thmstylethree}%

\begin{document}

\title[Violence Against Women: a pilot study on the perception of Apulian High school students]{Violence Against Women: a pilot study on the perception of Apulian High school students}

%%=============================================================%%
%% GivenName    -> \fnm{Joergen W.}
%% Particle -> \spfx{van der} -> surname prefix
%% FamilyName   -> \sur{Ploeg}
%% Suffix   -> \sfx{IV}

\author*[1]{\fnm{Crescenza} \sur{Calculli}}\email{crescenza.calculli@uniba.it}

\author[2]{\fnm{Serena} \sur{Arima}}%\email{serena.arima@unisalento.it}

\author[1]{\fnm{Alessio} \sur{Pollice}}%\email{alessio.pollice@uniba.it}

\author[1]{\fnm{Nunziata} \sur{Ribecco}}%\email{nunziata.ribecco@uniba.it}

\affil[1]{\orgdiv{Department of Economics and Finance},
\orgname{University of Bari Aldo Moro}, \city{Bari},
\country{Italy}}

\affil[2]{\orgdiv{Department of Human and Social Sciences},
\orgname{University of Salento}, \city{Lecce}, \country{Italy}}


\abstract{Violence Against Women (VAW) is a widespread issue deeply
rooted in social and cultural structures. Affecting women of all
ages and backgrounds, VAW is often underreported due to stigma and
victim-blaming. This study explores young people's perceptions of
VAW in the Apulia region (Southern Italy), using a local survey
inspired by a National framework on gender stereotypes and attitudes
towards VAW. The survey gathers insights into youth opinions on
gender roles, the acceptability of violence, and awareness of VAW
within their communities, aiming to uncover the underlying attitudes
that perpetuate this issue. The analysis combines two methodological
approaches to examine these data. A network-based approach explores
relationships within item responses, allowing for an in-depth look
at the direct interactions among youth attitudes. This approach is
paired with a psychometric model based on Item Response Theory,
specifically the Graded Response Model, which interprets attitudes
as manifestations of latent traits, revealing how different factors
shape perceptions of VAW.  Together, these methods offer a
comprehensive analysis of young people's views on VAW, highlighting
both individual response patterns and broader cultural trends
essential for designing effective interventions. Findings indicate a
gradual shift in attitudes toward gender roles; however, traditional
views remain prevalent, especially among young males. Socioeconomic
factors, such as parents' employment status, also contribute to the
persistence of stereotypes, underscoring the need for targeted
interventions to address and reduce VAW in youth populations.}




\maketitle

\section{Introduction}\label{Intro}

Violence against women (VAW) is a global public health issue
\citep{Stockl2024} strictly associated with social and cultural
structures. It refers to acts of gender-based violence (GBV) that
disproportionally affects women of all ages, racial, cultural and
economic backgrounds and is one of the most pervasive human rights
violations \citep{UN93}. The World Health Organization (WHO)
estimates that 30\% of women worldwide have experienced physical
and/or sexual violence from an intimate partner, or sexual violence
from a non-partner, at some point in their lives \citep{WHO2021}.
The phenomenon has multiple causes deeply rooted in gender
inequality \citep{Lomazzi2023}, power imbalances \citep{Fana2022},
and detrimental stereotypes \citep{Priyashantha2023, Stewart2021}.
These factors not only perpetuate violence \citep{Musso2020} but
also result in significant consequences, including physical and
mental health issues, economic constraints, and the perpetuation of
a poverty cycle that is challenging to disrupt \citep{Lausi2024,
Vyas2023, Sian2017}. One of the primary obstacles in addressing VAW
is the associated social stigma, coupled with the tendency to
attribute blame to victims \citep{Meluzzi2021} and the
underreporting of incidents leading to an underestimation of the
phenomenon \citep{Polettini2023}. In recent years, efforts to
highlight the significance of implementing precise measures for
quantifying VAW have been made \citep{Bettio2020}. However, data
sources remain scattered and inadequately updated, despite
legislation such as Law No. 53/2022, titled ``Provisions on
Statistics on Gender-Based Violence'' \citep{GazzettaUfficiale2022}
which underscores the importance of ensuring effective monitoring of
the phenomenon to develop adequate policies for the prevention and
tackling of GBV. Specifically in Italy, the most recent specialized
study on violence was conducted in 2014 by the National Statistics
Institute (ISTAT), titled ``Survey on the Safety of Women''
\citep{ISTAT2015}. On the other hand, a valuable national resource
for data on this issue is the official crime statistics register,
which is managed by the Italian Ministry of Interior and published
yearly \citep{MinIN2024}. While this register is crucial for
providing an overview of VAW-related crimes, it faces quality
issues, especially due to underreporting. Other data sources come
from local AVCs (Anti-Violence Centers), which play a crucial role
in supporting victims by offering essential supports (shelter,
counseling, legal assistance, etc.). These centers, established with
the Law No. 119/2013 \citep{GazzettaUfficiale2013}, act as gathering
information service on the prevalence and nature of gender-based
violence by systematically documenting instances of abuse, providing
valuable insights into the demographics of victims and the types of
violence experienced \citep{Toffanin2020}. However, privacy
requirements for accessing this data limit its immediate
accessibility and complicate its integration with other official
data sources.

While quantifying the phenomenon helps illustrate the scope and
scale of VAW, exploring its causes provides insights into the
social, cultural, and economic drivers that perpetuate such
violence. Studies indicate that factors such as labour market
participation and socioeconomic status act as drivers on attitudes
toward domestic and private violence \citep{Kuskoff2022,
Leguizamon2020, Armstead2021}. Also, educational attainment
\citep{Yoon2020, Kiss2012} and age \citep{Yoon2020} significantly
shape individuals' perceptions of VAW, which affects propensity of
reporting abuses \citep{Gracia2020}. Understanding these drivers is
essential for developing effective prevention and intervention
strategies \citep{Sheppard2024}. By identifying root causes,
researchers and policymakers can design targeted initiatives to
address and mitigate these conditions, reducing the incidence of VAW
while fostering a cultural shift toward gender equality, respect,
and healthier communities \citep{WHO2021, WHO2006}.

Young people play a pivotal role as agents of change in this
context. Adolescence and young adulthood (ages 10--24) are marked by
heightened expectations to conform to socially constructed norms,
which often perpetuate gender inequalities. The perspectives of
young people on VAW are shaped by a complex interplay of societal,
cultural, and educational influences thus understanding how these
elements impact their attitudes is crucial, as such beliefs either
reinforce or challenge existing norms related to gender inequality
and violence \citep{Kagesten2016}.

To this end, in this paper, we propose to investigate the attitudes
and opinions of young people regarding VAW through a local survey
conducted among youths in the Apulia region in Southern Italy. We
conducted a pilot study to gain exploratory insights into young
people's attitudes towards VAW in a specific local context. By
involving high-school students in the analysis, we aim to
investigate young people's perspectives on GBV, examining their
beliefs, attitudes, and awareness of VAW in their communities.
Inspired by the latest ISTAT specialized survey ``Stereotypes about
gender roles and the social image of sexual violence''
\citep{ISTAT2023}, our study adapts its framework to suit a younger
demographic. This tailored survey tool is designed to analyze
cultural patterns and potential factors related to local backgrounds
that may influence youth attitudes toward VAW. The used tool
includes questions assessing stereotypes about gender roles, views
on the acceptability of violence, its causes, as well as attitudes
toward sexual violence. It is worth highlighting that we consider a
sample of respondents from a specific territory in Southern Italy,
an area historically characterized by a socio-economic and cultural
divide compared to the North \citep{Salvati2017}. Territorial
inequalities are traditionally attributed to several factors,
including development, employment, and economic performance, as well
as differences in family and society structures \citep{Aassve2021}.
In Southern Italy, larger families and traditional roles within a
patriarchal framework may influence social dynamics and economic
opportunities, contrasting with the more prevalent nuclear families
in the North \citep{Caltabiano2019}. These factors can reinforce
traditional gender roles and limit opportunities for women,
perpetuating inequalities and affecting educational attainment, job
mobility, thereby contributing to the socio-economic contrast
between the two areas.

Different methodologies are used to analyze preference data obtained
from the administered survey. Preference data, such as rankings and
ratings, are prevalent in the social and behavioral sciences for
expressing and measuring attitudes or opinions \citep{Borsboom2015}.
To provide a more comprehensive understanding of trends and patterns
in young people's attitudes toward VAW, we propose a preliminary
approach that utilizes graphical tools to visually represent the
data, complemented by network analysis for a more in-depth
investigation of the relationships within the data. This
network-based approach aligns with recent psychometric
methodologies, wherein item responses are treated as proxies for
variables that interact directly with one another
\citep{Borsboom2021}. In these models, the covariance between
observable variables is interpreted as arising from direct
interactions among the variables themselves. This provides an
alternative to classical latent variable models, which treat
observables as manifestations of underlying factors
\citep{Bartholomew2011}. As an example, latent traits can be
estimated with Hidden Markov models (HMMs; see
\citealp{Colombi2024}) or within the Structural Equation Modeling
(SEM) framework \citep{Giovanis2023}.

%This approach therefore serves as a descriptive
%tool to explore patterns of association and potential interrelations
%within the data.

A machine learning approach for estimating Markov Random Fields
(MRF) or undirected networks for ordinal data \citep{Marsman2025} is
here proposed to uncover the underlying dynamics of attitudes and
opinions toward VAW and identify key factors influencing these
perceptions among young people. As discussed above, since latent
constructs can be interpreted as emergent properties arising from
the interaction between observed variables \citep{vanBork2019},
clusters of highly connected items may be viewed as representing
specific attitudes or belief systems related to VAW. Using the
Bayesian framework for MRF estimation allows for the incorporation
of uncertainty of structure and the identification of preference
heterogeneity within the population, thereby enhancing the
robustness of the findings.

For this work, we suggest complementing network model analysis with
the traditional psychometric approach grounded in Item Response
Theory (IRT). The majority of theoretical advancements in IRT
originated from psychometrics and educational measurement
disciplines. Significant early contributions to this field were made
by \cite{Rasch1960}, \cite{Birnbaum1968}, \cite{Wright1979}, and
\cite{Lord1980}. A focus on Graded Response Models (GRMs) introduced
by \cite{Samejima:1969} has been used to analyze preference data on
VAW. The GRM is a family of latent trait models designed for
assessing responses across ordered polytomous categories, such as
those arising from Likert-scale \citep{Likert1932} attitude surveys
\citep{Samejima1997}. An application of GRM is provided by
\cite{DelSarto2022}, who identify latent dimensions of spare-time
use and profiles of Italian Millennials based on the ``Multipurpose
Survey on Households: Aspects of Daily Life'' sample survey
\citep{ISTAT2016}.

Under these models, latent constructs are conceptualized as
underlying dimensions that explain patterns of responses to survey
items related to beliefs, stereotypes, and awareness about VAW. The
GRM operationalizes these constructs by estimating individual latent
traits that represent the intensity or degree to which respondents
endorse particular attitudes or beliefs. Each construct corresponds
to a dimension captured by a set of related items designed to
measure specific aspects of attitudes toward VAW (i.e., acceptance
of violence, gender role stereotypes, or perceived causes of
violence).

Thus, while network models highlight direct variable interactions
and connections, GRM offers a solid framework for examining how
survey items correlate with underlying latent traits. To evaluate
the likelihood of specific responses based on an individual's latent
traits and item characteristics, Bayesian inferences on GRM for the
perception of VAW are provided by MCMC implementation through the
R-package NIMBLE \citep{NIMBLE2024, deValpine2017}.

By combining these two psychometric approaches, we aim to deepen the
exploratory analysis of attitudes towards VAW using data from a
local survey. Our goal is to capture both the complex relationships
among various attitudes and opinions of young people and the
fundamental latent constructs driving them. To this end, network
analysis uncovers direct, pairwise associations among observed
variables revealing which beliefs co-occur and how tightly they
cluster, while the Graded Response Model (GRM) provides precise,
item-level measurements by estimating difficulty and discrimination
parameters for each response category. Unlike traditional factor
analysis (which reduces covariation to a few factors without
modeling item-by-item dependencies), this dual approach lets us map
interdependencies among items and estimate individual latent traits
in one cohesive framework. Together, these well-established,
complementary methods yield a richer, more nuanced, and more robust
interpretation of the data, advancing our substantive understanding
of young people's attitudes toward VAW. Our study thus demonstrates
the practical value and explanatory power of applying network
analysis and GRM in combination to real-world survey data.

%By combining these two psychometric approaches, we can deepen the
%analysis of attitudes towards VAW using data from local survey by
%capturing both the complex relationships among various attitudes and
%opinions of young people and the fundamental constructs driving
%them.

The structure of this article is as follows: Section~\ref{DataC}
introduces the case study, describing the local survey and the
sample characteristics. Section~\ref{Explo1} presents key findings
from the exploratory analysis, while Section~\ref{Explo2} discusses
the network-based approach used for analyzing collected item
responses. In Section~\ref{IRT}, we introduce the Item Response
Theory (IRT) framework, with a specific focus on the Graded Response
Model (GRM) in Section~\ref{IRT1}. The application of the GRM to
assess perceptions of VAW is outlined in Section~\ref{IRT2}.
Finally, concluding remarks and further developments are provided in
Section~\ref{Concl}.

\section{Data collection}\label{DataC}

To investigate the perception and the opinions about VAW among young
people, an online survey based on an tailored questionnaire was
conducted in February 2024. The questionnaire was primarily
developed by the research team at the University of Bari, Italy.
However, selected high school students involved in the Piano Lauree
Scientifiche - Statistica (PLS - Statistica) project
 sponsored by the Italian Ministry of Education and Merit \citep{PLS-MUR},
in collaboration with their teachers, contributed by suggesting
items, drafting questions, and providing feedback under close
supervision. This participatory approach aimed to ensure the
questionnaire was accessible and relevant to the target population.
The survey was administered to students from five high schools
participating in the PLS - Statistica project. Teachers distributed
the survey link, created using Google Forms within the Google
Workspace platform \citep{gsuite}, and students were invited to
complete it voluntarily. Anonymity and data protection were
guaranteed through formal acceptance of privacy policies. To expand
the sample, respondents were encouraged to share the survey link
with peers from their own or other schools. It is worth noting that
while data collection involved only high school students as
respondents, all data processing and analysis were conducted by the
University research team. The sampling and survey implementation
were intentionally designed to be accessible to students in the PLS
- Statistica project with limited statistical expertise, allowing
them to take part in the initial phases of the research process
while ensuring overall methodological soundness.

The resulting sample is a non-probability sample (convenience o
accidental sample) that proves to be highly effective for
exploratory, and qualitative studies \citep{Golzar2022}. The use of
non-random sampling does not aim to evaluate hypotheses about a
broader population, but rather to gain initial insights into a small
or understudied group, such as students in a specific territory
\citep{Peterson2014}. Furthermore, it offers the benefits of swift
and cost-effective data-gathering \citep{Etikan2015}.  The final
survey tool is an adjusted and customized version of the
questionnaire proposed by the National Institute of Statistics
(ISTAT) for the ``Survey on Attitudes, Perceptions, and Opinions
Regarding Gender Differences and the Social Image of Violence''
conducted in 2018 and 2023 \citep{ISTAT2023}. While the ISTAT survey
targeted the general population, the questionnaire used for this
study has been specifically tailored to suit a younger range of
respondents. A section addressing the personal and socio-demographic
characteristics of young respondents includes variables such as
gender (\emph{G}), age group (\emph{AG}), parents' educational level
(\emph{EduF} for father and \emph{EduM} for mother), and their
employment status (\emph{JbF} for father and \emph{JbM} for mother).
Along with this section, the questionnaire comprises 36 questions
with a total of 97 items designed to address different aspects of
opinions regarding gender-based violence among young people. These
aspects include: 1. Stereotypes about gender roles; 2 Stereotypes
about sexual violence; 3. Opinions on relationship dynamics; 4.
Behaviors that characterize toxic relationship. Furthermore, a
specific section of the questionnaire is dedicated to examining
experiences of violence, both direct and indirect, as well as the
processes of incident reporting. This focus facilitates a more
comprehensive understanding of the challenges individuals encounter
in reporting and the potential factors that influence their
decisions to disclose such incidents.

The entire questionnaire is designed with closed-ended questions
and, for the section on opinions about VAW, a 4-point Likert scale
is employed as a rating system \citep{Likert1932}. This approach
forces respondents to make choices, thereby preventing neutral
responses (\textit{e.g.}, ``1'' = \emph{fully disagree}, ``2'' =
\emph{disagree}, ``3'' = \emph{agree}, ``4'' = \emph{fully agree})
\citep{Capecchi2024}.

The total number of collected responses is 1,419 which dropped to
1,395 after data cleaning and consistency checks. All variables in
the final dataset are categorical or ordinal variables as derived
from Likert-type questions.

\begin{figure}[t]
\centering
\includegraphics[width=\textwidth]{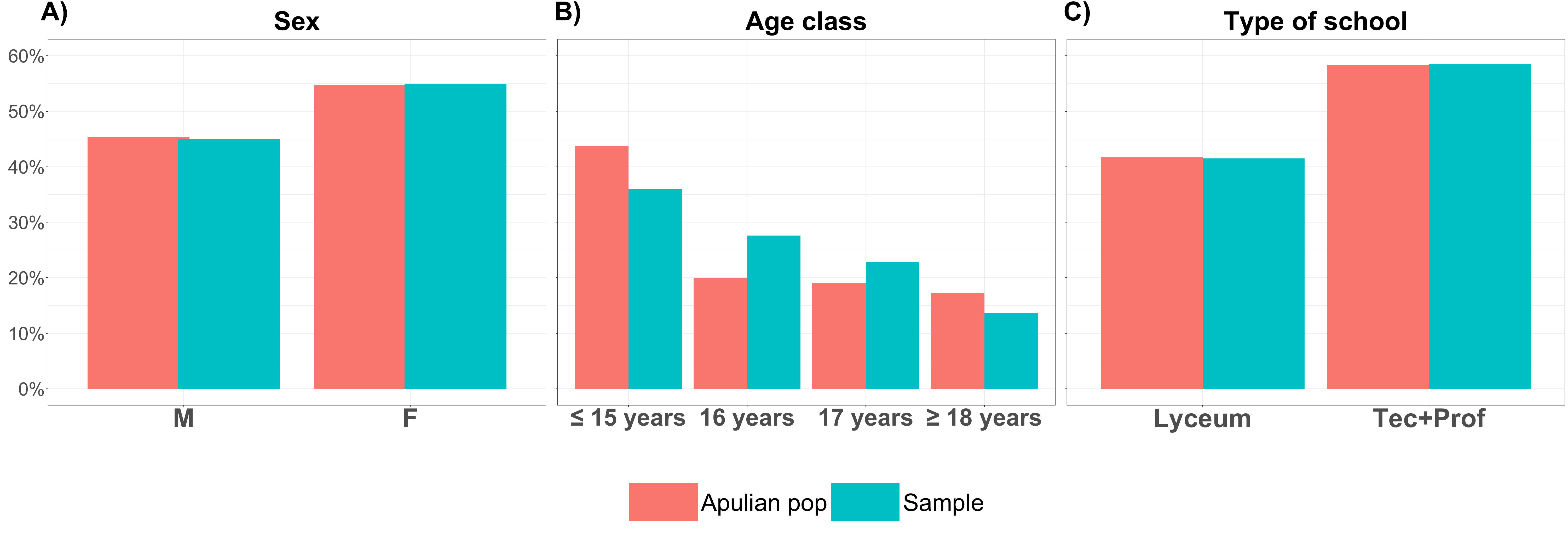}
\caption{Comparison between the distribution of respondents (Sample)
and the distribution of students at the regional level (Apulian
pop), based on sex, A), age class, B), and type of school attended,
C). For A), sex categories include male (M) and female (F) students.
In B), age classes include students aged up to 15 years, 16 years,
17 years, and 18 years and older. For C), types of school correspond
to the educational tracks attended, where ``Lyceum'' refers to
academic-oriented curricula (classical and scientific high schools)
and ``Tec+Prof'' stands for technical and professional
curricula.}\label{Fig1}
\end{figure}

\begin{figure}[t]
\centering
\includegraphics[width=0.9\textwidth]{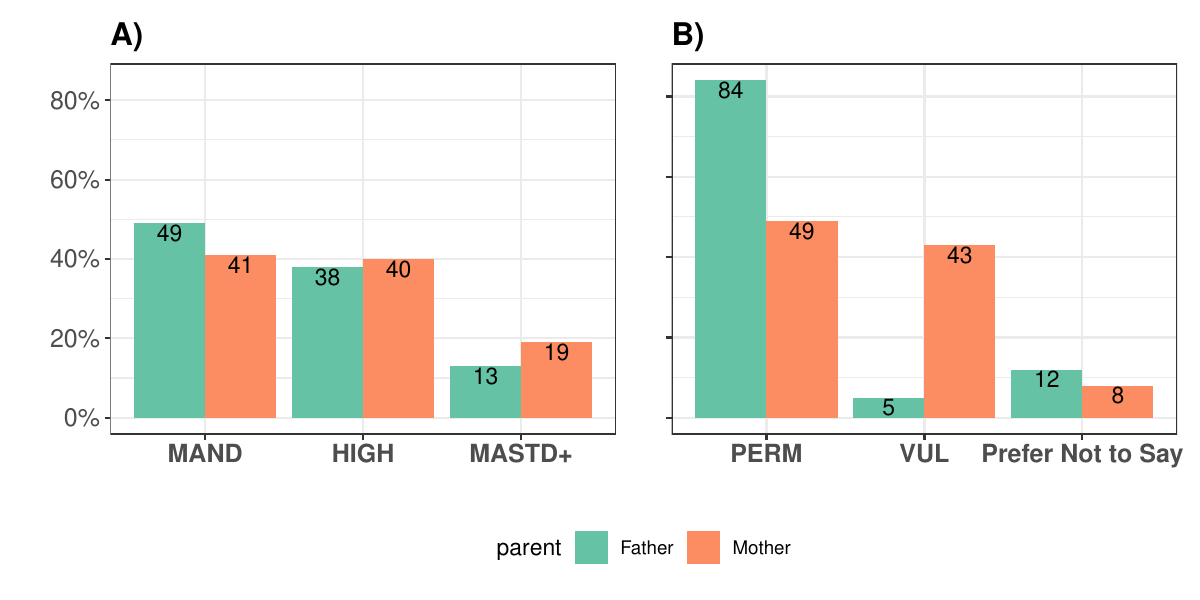}
\caption{Distribution of respondents based on the educational
levels, A), and employment statuses, B), of parents. For A),
abbreviations refer to the parents' level of education: MAND =
Mandatory school, HIGH = High school, MASTD = Bachelor's/Matser's
degree or higher. For B), abbreviations refer to the parents'
employment status: VUL = Vulnerable condition (precarious work,
on-call work, etc.), PERM = stable and permanent work.}\label{Fig2}
\end{figure}

The respondents' demographic profile reveals a higher proportion of
females (55\%) compared to males (45\%). Half of these fall within
the 16--17 age range, while the remaining individuals are split
between the 13 --15  (36\%) and the 18 -- 20 (14\%) age classes. The
distribution of respondents by type of school reveals that 41.5\% of
students attend lyceums, which include academic-oriented curricula
(classical and scientific high schools), while 58.5\% are enrolled
in technical and professional schools, covering technical institutes
and vocational programs. To ensure the sample of respondents matches
the main characteristics of the regional student population, we also
compared these key demographic factors using data provided by the
Italian Ministry of Education and Merit (retrieved from
\url{https://dati.istruzione.it/opendata/opendata/}).
Figure~\ref{Fig1} shows the distributions of sex, age class, and
type of school attended for both the sample and the regional student
population. The comparison suggests a strong alignment between the
two groups, indicating that the surveyed individuals closely reflect
the broader population in these respects. Considering other sample
characteristics, the majority of respondents (62\%) originates from
small towns and villages outside the Bari metropolitan area. With
regard to family background, the distribution of respondents by the
educational level and the employment status of parents, indicates
that mothers possess a higher level of education compared to
fathers, albeit with more precarious working conditions, as
illustrated in Figure~\ref{Fig2}.

\section{Some results from exploratory tools}\label{Explo}

\subsection{Descriptive results}\label{Explo1}

\begin{table}[t]
\caption{Trigger items on 4-point Likert agreement scale with ``1''
= \textit{fully disagree} and  ``4'' = \textit{fully agree}.
``Abbr.'' are abbreviations used to identify items}\label{Tab1}%
\begin{tabular}{llll}
\toprule
 &   Section                                              & Item                                                           & Abbr.\\
 \midrule
 & \multirow{8}{*}{\footnotesize 1. Stereotypes about gender roles}     & \footnotesize \multirow{2}{*}{Mother's responsability to look after children} & \multirow{2}{*}{MRC}\\[3pt]
 &                                                        & \footnotesize and take care of their daily needs                             & \\
 &                                                        & \footnotesize \multirow{2}{*}{A good wife/partner should comply with her }   & \multirow{2}{*}{CHI}\\[3pt]
 &                                                        & \footnotesize husband's/partner's ideas even if she disagrees                &  \\
 &                                                        & \footnotesize \multirow{2}{*}{Men should provide for the economic }          & \multirow{2}{*}{MNF}\\[3pt]
 &                                                        & \footnotesize needs of the family                                            &\\
 &                                                        & \footnotesize \multirow{2}{*}{Employers should give preference to }          & \multirow{2}{*}{PrM}\\[3pt]
 &                                                        & \footnotesize men over women                                                 &\\
\midrule
 & \multirow{8}{*}{\footnotesize 2. Stereotypes about sexual violence}  & \footnotesize \multirow{2}{*}{If a woman is raped after accepting} & \multirow{2}{*}{AIR}\\[4pt]
 &                                                        &   \footnotesize a man's invitation, she is also at fault                         &\\
 &                                                        & \footnotesize \multirow{2}{*}{Women often say no but actually mean yes}          & \multirow{2}{*}{NmY}\\[3pt]
 &                                                        &                                                                                  &\\
 &                                                        & \footnotesize \multirow{2}{*}{Women can provoke sexual violence by }             &  \multirow{2}{*}{DrE}\\[3pt]
 &                                                        & \footnotesize the way they dress   &\\
 &                                                        & \footnotesize \multirow{2}{*}{Forcing a wife/partner to have non-consensual}     &  \multirow{2}{*}{IPV}\\[3pt]
 &                                                        & \footnotesize act is not violence    &\\
\midrule
 & \multirow{8}{*}{\footnotesize 3.  Opinions on relationship dynamics} & \footnotesize Betrayal is always morally unacceptable   &  BET\\
 &                                                        &              &\\
 &                                                        & \footnotesize \multirow{2}{*}{Female infidelity is always more serious }      &\multirow{2}{*}{FIN}\\[3pt]
 &                                                        & \footnotesize than male infidelity    &\\
 &                                                        & \footnotesize \multirow{2}{*}{Male infidelity is often justifiable}           &\multirow{2}{*}{JMI}\\
 &                                                        &                                                    &\\
 &                                                        & \footnotesize \multirow{2}{*}{The partner is property of the mate even }        &\multirow{2}{*}{PrP}\\[3pt]
 &                                                        & \footnotesize after the relationship ends       &\\
 \midrule
 & \multirow{7}{*}{\footnotesize 4.  Toxic behaviors}     & \footnotesize Continuous  messages & CoM\\
 &                                                        &                                                 &    \\
 &                                                        & \footnotesize Check partner's phone             & CPP \\
 &                                                        &                                                 &    \\
 &                                                        & \footnotesize Distancing from family and friends & DFF \\
 &                                                        &              &\\
 &                                                        & \footnotesize Sharing a social media account     & SHS\\
% & \multirow{2}{*}{4. Direct and/or undirect experience of}  &
% Question \\
% &   violence and reporting                                 &       \\
 \bottomrule
\end{tabular}
\end{table}

\begin{figure}[t]
\centering
\includegraphics[scale=0.31]{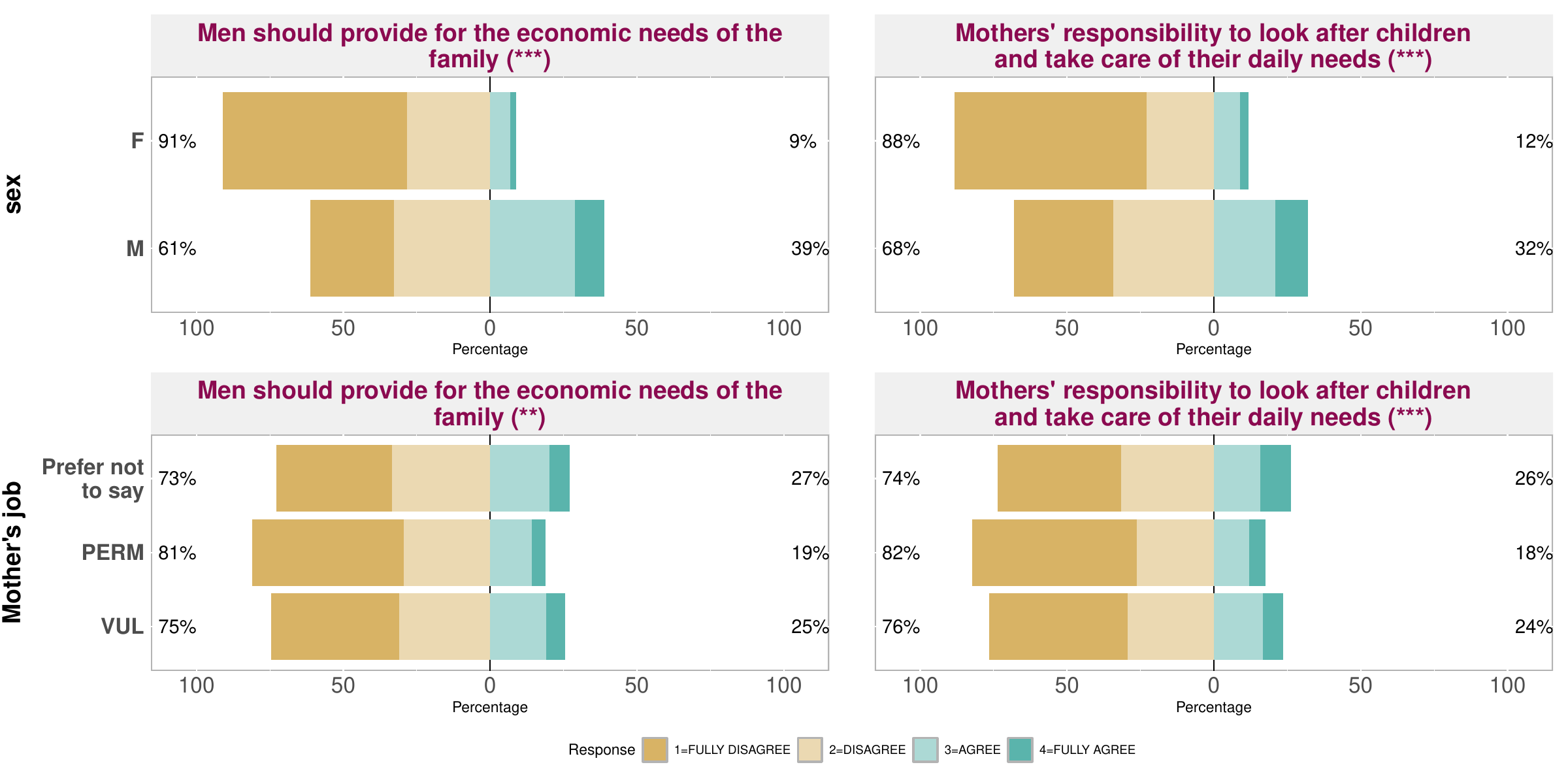}
\caption{Distribution of scores assigned to selected items
concerning stereotypes about gender roles. Based on Mood's median
test, (***) represents p-value$<0.001$ and (**) represents p-value$<
0.01$. Referring to the mother' employment status, abbreviations VUL
and PERM stand for ``Vulnerable work condition'' and  ``stable and
permanent work''.}\label{Fig:2}
\end{figure}

\begin{figure}[t]
\centering
\includegraphics[scale=0.31]{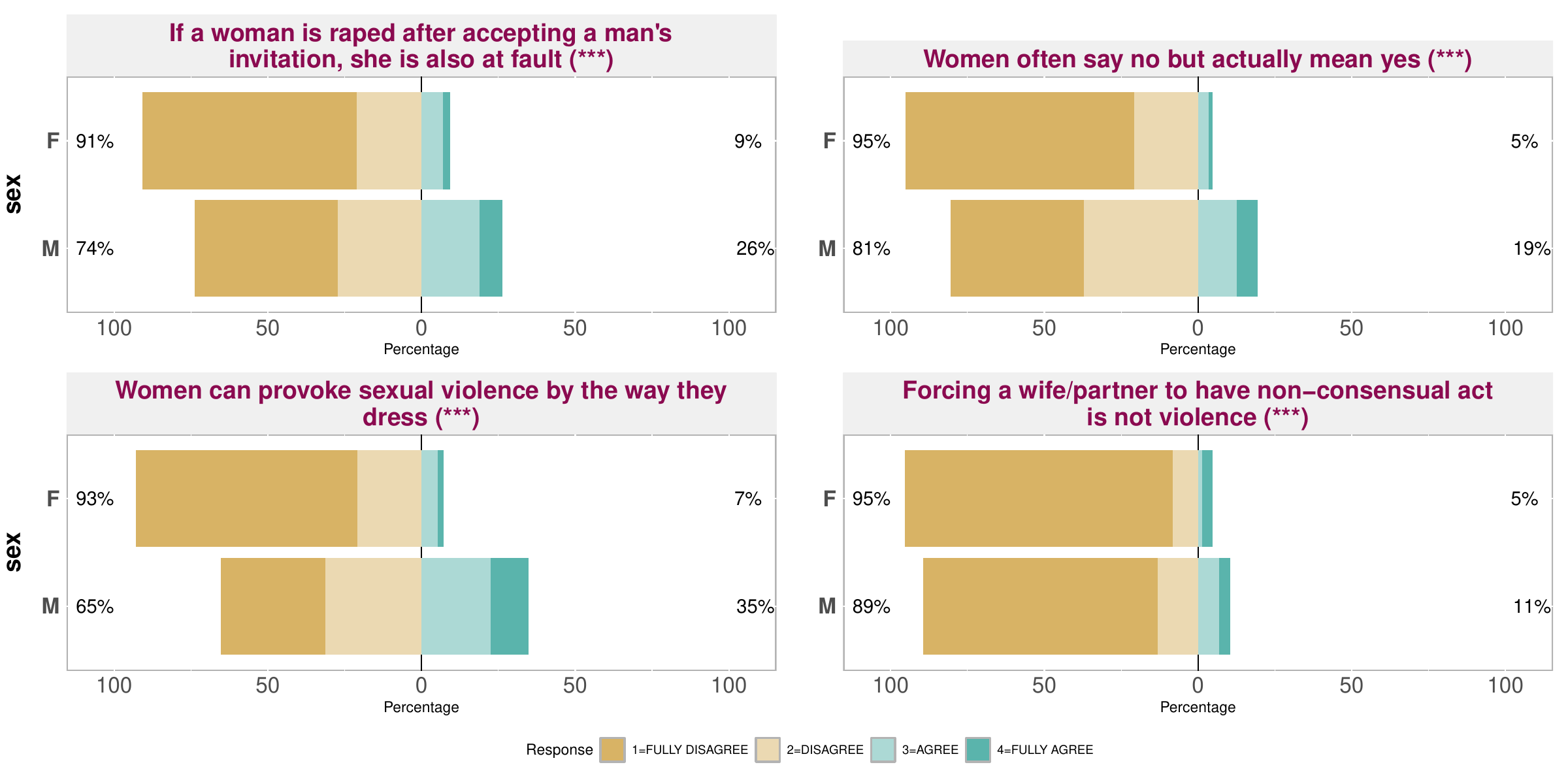}
\caption{Distribution of scores assigned to selected items
concerning stereotypes about sexual violence. Based on Mood's median
test, (***) represents p-value$<0.001$ and (**) represents p-value$<
0.01$. Referring to the mother' employment status, abbreviations VUL
and PERM stand for ``Vulnerable work condition'' and  ``stable and
permanent work''.}\label{Fig:3}
\end{figure}

\begin{figure}[t]
\centering
\includegraphics[scale=0.5]{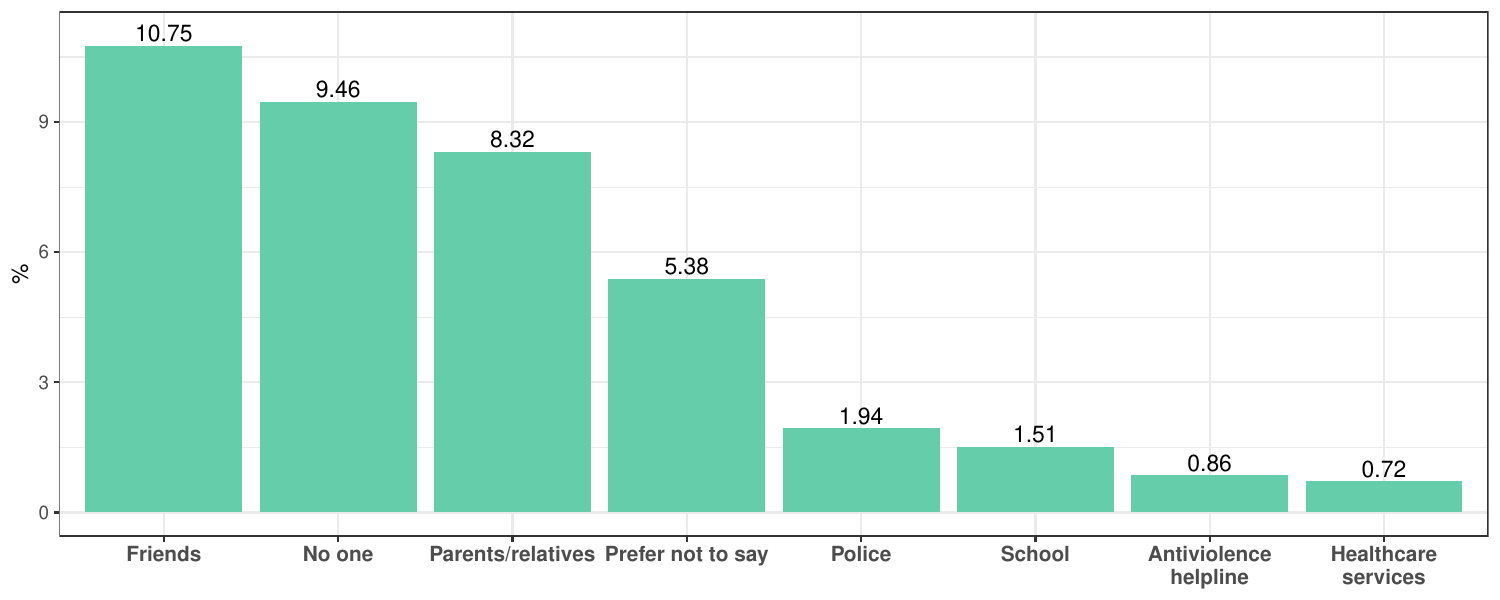}
\caption{Incidents of violence reported by young respondents by type
of reporting.}\label{Fig:4}
\end{figure}

A subset of trigger questions and items, representative of each
investigated aspect of the phenomenon, were selected for the
exploratory data analysis and is reported schematically in
Table~\ref{Tab1}. The distribution of scores assigned by respondents
to the selected items is preliminarily investigated in relation to
potential influencing variables using graphical tools and
statistical tests. Given the ordinal nature of the data and the
presence of skewed distributions, the Mood's median test
\citep{Mood:1954} was employed to compare item scores across
subgroups. These subgroups were defined by key demographic and
socioeconomic characteristics (i.e., gender, age group, education
level, and parents' employment status) to assess whether these
factors were associated with systematically higher or lower
evaluations. To facilitate this investigation, visual
representations are created using the \texttt{likert} package in
\texttt{R} software \citep{likertpack2016}. For the sake of brevity,
only a subset of these items is presented, focusing on the effect of
a few factors in shaping stereotypes. The results in
Figure~\ref{Fig:2} reveal notable disparities in the perception of
gender roles across demographic groups, with gender emerging as a
particularly strong influence. Males, in particular, are more likely
to endorse traditional social stereotypes related to work, household
responsibilities, and child-care duties. These stereotypes are also
more prevalent among respondents who report their mother's
employment status as vulnerable or who prefer not to disclose it.
Exploring stereotypes about sexual violence, Figure~\ref{Fig:3}
shows the victim-blaming attitudes distinguishing between females
and males. Significant differences across genders are found in how
sexual violence is perceived and understood. The prejudice that
assigns responsibility to females who suffers sexual violence seems
to be widespread among males. This gender divide in victim-blaming
attitudes has far-reaching implications for how sexual violence
cases are addressed and how survivors are treated. The tendency to
hold female victims accountable for their experiences not only
perpetuates harmful stereotypes but also contributes to a culture of
silence and shame surrounding sexual violence. Such attitudes can
discourage reporting, hinder justice, and impede the healing process
for survivors. Following this consideration, a focus on the
experience of violence indicates potential misreporting among young
respondents who declare themselves as victims of abuse. In this
study, 4\% of the total respondents reported having been victims of
physical violence, while 6\% reported having experienced both verbal
and psychological violence. Regardless of the type of violence, as
reported in Figure~\ref{Fig:4}, approximately the 15\% of these
victims do not report incidents of abuse, despite having experienced
them (``No one'' and ``Prefer not to say'' choices). A small
percentage utilize anti-violence centers or other services, which
may evade monitoring and contribute to the underestimation of the
phenomenon.

\subsection{Network analysis}\label{Explo2}

According to a more recent psychometric approach for investigating
preference data \citep{Borsboom2021, Marsman2018}, a more detailed
exploratory analysis based on a machine learning approach is
performed using a Markov Random Field (MRF) or undirected graphical
model \citep{Epskamp2022} to assess the underlying structure of
interconnections between items.

This type of model belongs to the broader class of graphical models
\citep{Maathuis2018, lauritzen1996}, where the network graph encodes
probabilistic relationships between random variables (\emph{i.e.},
nodes). In MRF's, the connections between nodes (\emph{i.e.}, edges)
are undirected, implying no assumption of causal directionality.
Specifically, the edges represent the conditional associations
between pairs of variables after controlling for all other variables
in the model. The two main aspects of interest in such models are
the network structure, $\mathcal{G}$ and the matrix of association
strengths, $\mathbf{\Theta}$. The network structure, $\mathcal{G}$,
is comprised of unweighted, undirected edges, where the presence of
an edge indicates conditional dependence, and the absence indicates
conditional independence. As the number of variables increases, the
potential number of network structures grows exponentially, making
it crucial to quantify the uncertainty associated with the structure
estimation. Once the structure is determined, attention shifts to
the strength and nature of the relationships between variables,
encoded in $\mathbf{\Theta}$. Positive values of $\theta_{ij}$
represent positive associations between variables $i$ and $j$, while
negative values reflect negative associations. The magnitude of
$\theta_{ij}$ indicates the strength of this association, with
larger absolute values implying stronger relationships.

Within the Bayesian framework, the posterior distribution in
Equation \ref{Eq:1} describes the network estimation learning
process that can be used to quantify the uncertainty associated with
both the structure and its parameters

\begin{equation}\label{Eq:1}
\underset{\mbox{\tiny joint posterior}}
{p(\Theta,\mathcal{G}|\mbox{data})} \propto \underset{\mbox{\tiny
likelihood}} {p(\mbox{data}| \Theta, \mathcal{G})} \times
\underset{\begin{subarray}{c}
  \mbox{\tiny prior on} \\
  \mbox{\tiny parameters}
  \end{subarray}}{p(\Theta|\mathcal{G})} \times \underset{\begin{subarray}{c}
  \mbox{\tiny prior on} \\
  \mbox{\tiny structure}
  \end{subarray}}{p(\mathcal{G})}
\end{equation}

\noindent To assess the connection between two nodes $i$ and $j$,
the two hypotheses ``$\mbox{H1}_{(ij)}$: There is an edge between
the variables $i$ and $j$'' and ``$\mbox{H0}_{(ij)}$: There is no
edge between the variables $i$ and $j$'', can be tested once
obtained the posterior edge inclusion (exclusion) probabilities as
the sum of the posterior probabilities of all structures
$\mathcal{G'}$ that include (exclude) the edge between the
variables. This method, known as Bayesian model-averaging (BMA)
\citep{Hinne2020}, treats each possible graph structure as a
distinct statistical model. Rather than selecting a single most
likely model, it aggregates across all possible models. Once
probabilities of inclusion (exclusion) are obtained, a measure that
allows to determine which hypothesis is more strongly supported by
the data, can be derived as follows

\begin{equation}
\mbox{BF}_{10} = \underset{\begin{subarray}{c}
  \mbox{\tiny posterior inclusion} \\
  \mbox{\tiny odds}
  \end{subarray}}{\frac{p(\mbox{H1}_{ij}|\mbox{data})}{p(\mbox{H0}_{ij}|\mbox{data})}}/
\underset{\begin{subarray}{c}
  \mbox{\tiny prior inclusion} \\
  \mbox{\tiny odds}
  \end{subarray}}
{\frac{p(\mbox{H1}_{ij})}{p(\mbox{H0}_{ij})}}
\end{equation}

The indicator $\mbox{BF}_{10}$ represents the inclusion Bayes
factor, serving as a measure of the strength of evidence in the data
supporting edge inclusion (conditional dependence) \citep{Huth2023,
Wagenmakers2016}. On the other hand,
$\mbox{BF}_{01}=1/\mbox{BF}_{10}$ quantifies the evidence for edge
exclusion (conditional independence). Therefore, a
$\mbox{BF}_{10}=10$ suggests that the data are ten times more likely
under a network structure that includes the edge between nodes $i$
and $j$ than under a structure that excludes that edge.

To quantify parameter uncertainty, the posterior distribution of the
partial association parameters $p(\bf \Theta | \mathcal{G}, \mbox
{data})$ can be used to obtain posterior standard deviations and
credible intervals for each partial association parameter
$\theta_{ij}$.

For the analysis of ordinal data reflecting the respondents' scoring
process of the VAW case study, Bayesian inference on the network
structure and parameters is conducted using a novel approach
proposed in \cite{bgms2023}. This method employs a two-component
mixture prior for edge weights: a diffuse prior (slab) for included
edges and a spike prior assigning a weight of zero for excluded
edges, determined by a latent indicator variable. For the slab
specification, a Cauchy distribution with a large scale of 2.5 is
considered \citep{Gelman2008}, and a Bernoulli prior with edge
inclusion probability of 0.5 is assigned to the latent indicator. To
address computational challenges, such as the complexity of
normalizing the MRF and handling different parameter dimensions, a
pseudolikelihood approach \citep{Liang2010} and a transdimensional
Markov chain method \citep{Gottardo2008} are utilized. These methods
are implemented through Gibbs sampling in the R-packages
\texttt{bgms} \citep{bgms2023} and \texttt{easybgm}
\citep{easybgm2024}. The latter R-package unifies key features from
three Bayesian graphical modeling tools (\texttt{bgms},
\texttt{BDgraph} \citep{Mohammadi2019} and \texttt{BGGM}
\citep{Williams2019}), providing a comprehensive solution for
analyzing various types of data in social science research.

\begin{figure}[t]
\centering
\includegraphics[scale=0.51]{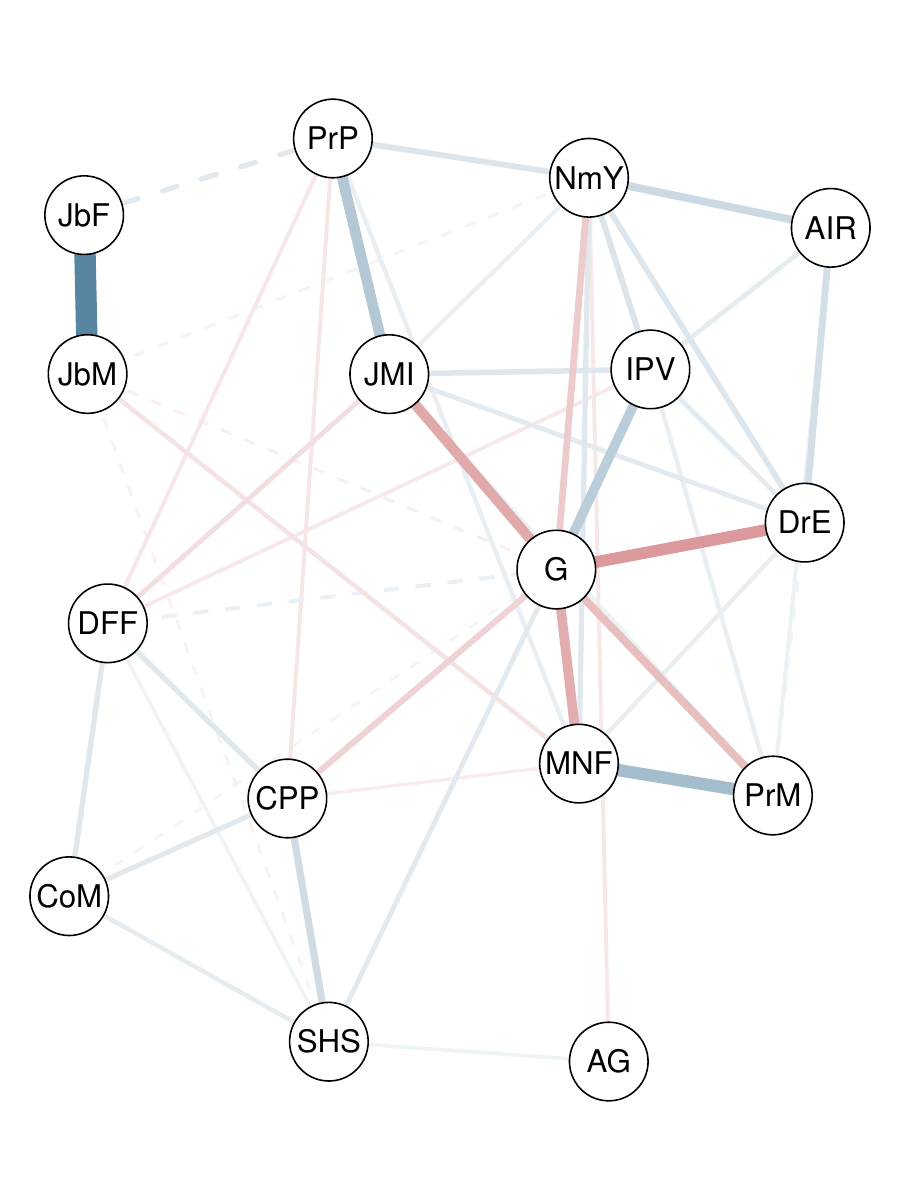}
\caption{The estimated MRF structure containing selected items and
individual characteristics from the VAW scoring assessment. Blue
edges represent positive correlations and red edges represent
negative correlations; the greater saturation of the edge indicates
a stronger correlation. Edges with insufficient evidence are shown
with a dashed line. Nodes abbreviations are reported in
Table~\ref{Tab1}.}\label{Fig:netw}
\end{figure}

The joint posterior distribution for the data from the VAW case
study was obtained by running 20,000 MCMC iterations with a burn-in
phase of 5,000. Figure~\ref{Fig:netw} shows the visualization of the
estimated MRF model, which includes selected items related to VAW as
well as relevant individual characteristics (e.g., gender, age group
and parent's employment status). For the sake of result clarity, not
all items in Table~\ref{Tab1} were included in the network
estimation, as some were considered redundant. As is common in
psychometric networks, a dense structure is observed, with all nodes
connected to at least one other node.  The network is shown by a
so-called median probability plot \citep{Barbieri2020}, which
excludes all edges with a posterior inclusion probability less than
0.5. Edges with an $\mbox{BF}_{10}<10$ are considered inconclusive
and shown with a dashed line. The estimated network suggests
positive correlations among items belonging to the same section of
the questionnaire. In this network framework, fundamental constructs
are identified based on clusters of nodes with strong
interconnections, representing groups of items that collectively
capture specific dimensions of attitudes or beliefs towards VAW. For
instance, items related to toxic behaviors (i.e., nodes CoM, CPP,
DFF, and SHS) and stereotypes about sexual violence (i.e., nodes
AIR, NmY, DrE, and IPV) are densely interconnected, forming clusters
of variables that are strongly related to one another, reflecting
similar patterns in the scoring process. The gender (\emph{G}) also
plays a clear role into patterns of responses. Females are
associated with lower scores concerning all stereotypes. They are
less likely to agree with statements regarding gender stereotypes,
such as the belief that men should provide for the economic needs of
the family (node MNF) or that employers should give preference to
men over women (node PrM). Females are also less likely than males
to agree with stereotypes regarding sexual violence. For instance,
they reject the belief that it can be provoked by a person's attire
(DrE) and tend to view non-consensual relationships within a couple
as a more serious issue (positive correlation with node IPV).
Furthermore, the age variable (AG) does not appear to influence
scores, despite that older age groups tend to demonstrate less
agreement with stereotypes concerning men's role as the primary
provider for the family's economic needs (MNF). A marginal effect of
variables connected to the employment status of parents (JbF and
JbM) is estimated in the network. These nodes exhibit a strong
positive correlation, suggesting that parents typically share
similar employment statuses within the family background. However,
only the mother's work situation appears to be associated to gender
role stereotypes. Specifically, mothers with more stable employment
are less likely to endorse the notion that men should be the primary
economic providers for the family. Another notable finding is the
negative relationship between the characterization of toxic
relationships and perceptions of relationship dynamics. It is
unsurprising that individuals who view a partner as `property,' even
after a relationship has ended, tend to assign lower scores when
assessing the toxicity of behaviors such as examining a partner's
phone (CPP) or isolating them from their social circle (DFF).
While these findings demonstrate that the network-based approach
contributes to the initial understanding of the complex
relationships and interactions among variables in the context of
gender-based violence, the next section will examine how Item
Response Theory (IRT) facilitates the analysis of individual traits
in relation to assessment performance and evaluates the measurement
properties of test items.

\section{Some results from Item Response Theory}\label{IRT}

\subsection{Introduction to IRT models}

Psychologists often measure latent traits such as intelligence,
anxiety, depression, or personality, through questionnaires and
standardized tests. Since these traits cannot be observed directly,
they are inferred from patterns of responses to items specifically
designed to reflect them. Psychometrics is the field of research
concerned with the theory and techniques of psychological
measurement. Psychological measurement has evolved from Classical
Test Theory (CTT) to more sophisticated, mathematically grounded
approaches \citep{Frey2017, Feldt1989, Lord1968}. CTT emerged in the
early 20\textsuperscript{th} century and focused on total test
scores, assuming that every test-taker's observed score is composed
of a true score plus error. Despite its widespread use, CTT has
notable limitations. For example, item statistics (such as
difficulty) are sample-dependent, and test scores are not invariant
across different populations (see, for instance, the foundational
work by \cite{Hambleton1991}). In 1960, Rasch introduced the Rasch
Model (1PL) \citep{Rasch1960}, which models the probability of a
correct response using only the item difficulty parameter. Rasch
emphasized the principle of specific objectivity, namely that
comparisons between individuals ought to be independent of the
particular items employed. The 1PL model was later extended to
include additional item parameters: the 2PL model, which adds a
discrimination parameter, and the 3PL model, which also includes a
guessing parameter (see \cite{vanderLinden1997} for a comprehensive
review). These models describe, with different parameterizations,
the probability of a correct response to an item as function of
test-taker's latent trait level: such probabilities are represented
by a curve named Item Characteristic Curve (ICC). ICCs of the Rasch
model are parallel to one another since they only depend on a single
item parameter (item difficulty). The ICCs of the 2PL may on the
other hand cross each other: the location of the ICC is determined
by the difficulty parameter, while the slope is defined by the
discrimination parameter (see \cite{Boomsma:1992} for a complete
review). While IRT models were initially developed for dichotomous
items, they have since been extended to accommodate polytomous
items, that is, those with more than two response categories. This
extension is crucial for modeling data from Likert scales, partial
credit scoring, and graded responses, which are commonly used in
psychological and educational assessments. In this context, we focus
on the so-called Graded Response Model (GRM) introduced in
\cite{Samejima:1969}, a global logit model \citep{Agresti:2002}
particularly suited for ordered categorical data and widely applied
when the latent trait is assumed to be continuous; this model will
be presented in more detail in the following section.

\subsection{The Graded Response Model}\label{IRT1}

Let $Y_{ij}$ be the response of the subject $i$ to the $j-$th item
of the questionnaire, with $j=1,\ldots,M$ and $i=1,\ldots,n$. The
response variable is a categorical variable with $H$ possible
ordered categories. Let $\Theta$ denote the latent trait that the
test aims at measuring and denote the probability $p_{ijh}$ that the
subject $i$ with latent trait $\theta_i$ answers by category $h$  to
item $j$ as

\begin{eqnarray}
p_{ijh}&=&P(Y_{ij}=h|\Theta_i=\theta_i) =\\
& & P(Y_{ij} \geq h|\Theta_i=\theta_i) -
P(Y_{ij} \geq h+1 |\Theta_i=\theta_i)= \\
& & p^{+}_{ijh}(\theta_{i}) - p^{+}_{ijh+1}(\theta_{i})
\hspace{1cm} h=1,...,H-1
\label{prob}
\end{eqnarray}

The probability that subject $i$ with latent trait $\theta_{i}$ answers with category $h$ or higher than $h$ to item $j$ is modelled as

\begin{equation}
{\rm logit}(p^{+}_{ijh})=\log \left(\frac{P(Y_{ij}\geq h|\theta)}{P(Y_{ij} < h|\theta)}\right)=\gamma_j (\theta_i - \beta_{jh}) \hspace{1cm} h=1,...,H-1
\label{GRM}
\end{equation}
where $\gamma_j$ defines the discrimination power of the item $j$
and $\beta_{jh}$ is the difficulty of response category $h$ of the
item $j$. The Graded Response Model is therefore a nonlinear mixed
model in which the probability  that a subject $i$ responds with
category $h$ or higher to item $j$ is modelled as a function of both
subject latent trait and item characteristics, i.e., difficulty and
discrimination power \citep{Rijmen:2003}. The difficulty parameter
is specified as

\begin{equation}
\beta_{jh}=\beta_j + \delta_{h} \hspace{1cm} j=1,...,M,
\hspace{0.5cm} h=1,\ldots,H-1 \label{difficulty}
\end{equation}

where $\beta_j$ denotes the difficulty of item $j$ and $\delta_{h}$
is the difficulty of response category $h$ for all items.
The discrimination parameter is constrained to assume positive values  as the probability of answering correctly is deemed
 to be increasing as an examinee's latent trait increases.

In what follows, we adopt a Bayesian approach according to
\cite{Fox:2013}
thus the model can be written as a multilevel model:

\begin{equation}
\label{eq:01}
logit(P(Y_{ij} \geq h| \theta, \gamma, \beta))=\gamma_j (\theta_i - \beta_{jh})
\end{equation}
\begin{equation}
\label{eq:03}
\beta_{jh} = \beta_{j}+\delta_{h}
\end{equation}
\begin{equation}
\label{eq:02}
\theta_i| \gamma, \beta \sim N(x_i^{T}\alpha,\sigma^{2}_{theta})
\end{equation}
\begin{equation}
\label{eq:04}
{\beta_j \choose log(\gamma_j)} \sim  N(\mu_{j}, \Sigma=diag(\sigma^{2}_{\mu}))
\end{equation}
\begin{equation}
\label{eq:05}
\delta_{h} \sim N(0, \sigma^{2}_{h});
\end{equation}
\begin{equation}
\label{eq:06}
\alpha \sim N(\mu_{\alpha},\sigma^{2}_{\alpha})
\end{equation}

\subsection{GRM for the perception of VAW}\label{IRT2}

We model the responses to the questionaire ($J=$25 items) described
in Section~\ref{DataC} with the GRM in Equations (\ref{eq:03} -
\ref{eq:06}). For the GRM, we set $\sigma^{2}_\theta=1$, to ensure
model identifiability and
$\sigma_{\mu}=\sigma^{2}_{h}=\sigma^{2}_{\alpha}=0.01$, reflecting
vague prior information and $\mu_{j}=0$ ($j=1,\ldots,25$) As
explanatory variables of the mean latent trait
(Equation~\ref{eq:02}), we use the following variables: sex
(male$=1$, female$=0$), age (in years), mother's employment status
(unemployed$=1$, employed$=0$) and father's employment status.
Interactions have been considered but their posterior distributions
were centered around zero implying that they do not have a key role
in explaining the variability of the latent traits. We implement the
model in R package NIMBLE \citep{deValpine2017, NIMBLE2024}: we run
2 chains and we allow 15000 iterations with burn-in of 5000 and
thinning rate of 10. Convergence of the MCMC chain was assessed by
visual inspection of the time series plots of selected variables as
well as by inspecting Gelman-Rubin diagnostic. In this framework,
fundamental constructs are identified through the estimation of
individual latent traits \(\theta_i\), which represent the
respondents' position on the underlying attitude continuum toward
violence against women (VAW). The discrimination parameters
\(\gamma_j\) play a key role by indicating the effectiveness of each
item \(j\) in differentiating respondents along the latent trait.
Specifically, higher values of \(\gamma_j\) correspond to items that
better discriminate between different levels of the latent attitude,
thereby highlighting the core constructs captured by the
questionnaire. This allows for classification and interpretation of
constructs based on the items' sensitivity in measuring meaningful
variation in attitudes.

\begin{figure}[t]
\centering
\includegraphics[scale=0.35]{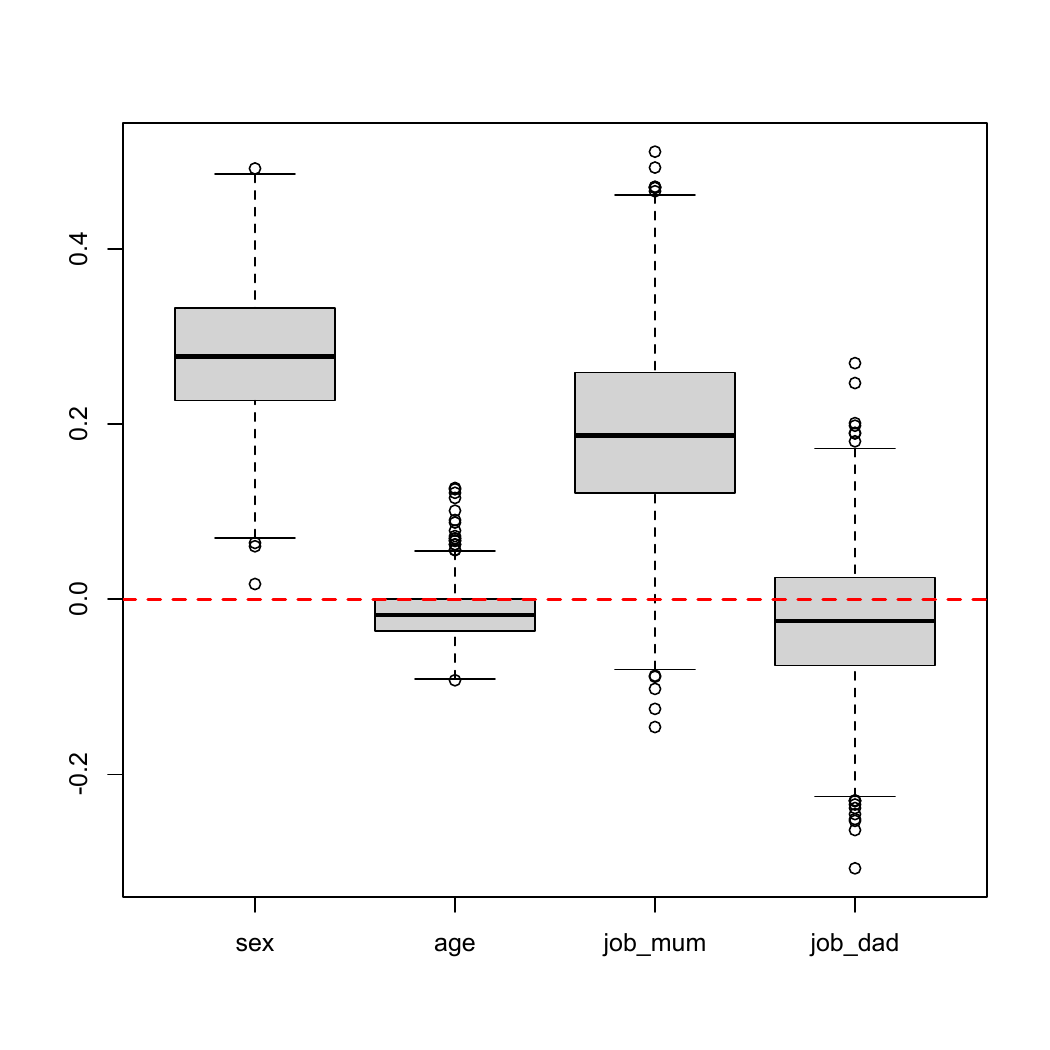}
\quad
\includegraphics[scale=0.35]{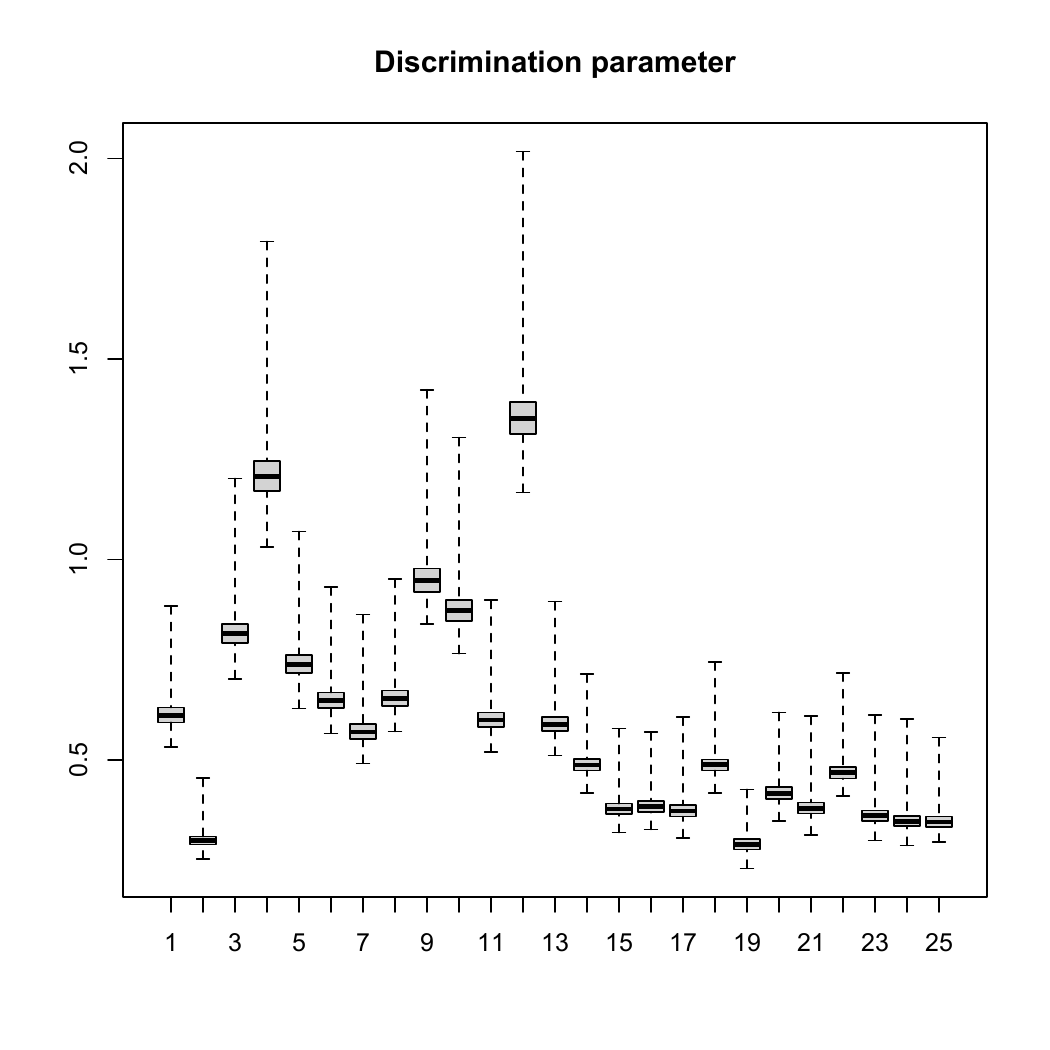}
\caption{Left panel: Posterior distribution of the GRM model
parameters. Right panel: posterior distribution of the item
discriminant parameters.}\label{PostDist}
\end{figure}

Left panel of Figure \ref{PostDist} shows the posterior
distributions of the model parameters: the gender and the mother's
employment status seem to play a role in influencing the gender
stereotypes perception. Males show a higher attitude to stereotypes
with respect to girls, strongly agreeing with questions highlighting
an asymmetric gender role. Similarly, students who declared that
mothers are employed are less likely to show agreement with
stereotypical assessments with respect to those whose mothers are
unemployed.

The right panel of Figure \ref{PostDist} shows the posterior distribution of the discriminant parameters:
items with the highest discriminant power regard the perception of the partner as a property.
The most discriminant items correspond to the following questions:

\begin{itemize}
\item The partner is property of the mate even after the relationship
ends;
\item If a husband/partner forces his wife/partner to have sexual
intercourse against her will, it is not violence;
\item Male infidelity is often very justifiable.
\end{itemize}

On the other hand, the following items are identified as the less
discriminant ones:

\begin{itemize}
\item Encouraging each other is dangerous in a toxic relation;
\item Women who do not want to have sexual intercourse are able
to avoid it.
\end{itemize}

\section{Conclusions}\label{Concl}

In this work, we present the results of a local study on attitudes
toward VAW, providing insights into some specific cultural, social,
and economic factors that contribute to this issue within a
particular community. The study is conceived as a preliminary,
context-specific investigation aimed at exploring young people's
attitudes toward VAW, with no claim to generalizability beyond the
boundaries of the case under study. For the analysis, we used two
complementary methodologies: an exploratory Network analysis and a
structured modeling approach through the Graded Response Model based
on Item Response Theory. The first approach investigates connections
among variables, revealing groups of attitudes associated with
gender stereotypes and detrimental behaviors. Concurrently, the
Graded Response Model offers a methodical framework for
understanding how individual responses relate to specific belief
patterns and the endorsement of stereotypes. Although these
techniques differ, they reveal a coherent picture of the attitudes
and perceptions shaping young people's views on VAW at the local
level. Convergent findings highlight that gender and maternal
employment status significantly influence perceptions of gender
stereotypes. Young males are still more likely to endorse
stereotypes and agree with asymmetric gender roles, while young
females tend to reject such beliefs, including the notion that men
should be the primary economic providers. The Network analysis
reveals that females tend generally to score lower on stereotypes
and perceive non-consensual relationships as more serious. These
results are encouraging and, although limited in terms of
generalizability, they provide a clearer picture of young people's
views on VAW. We believe that extending the study to the national
level would lead to a more comprehensive understanding of the
phenomenon. These findings also underscore the necessity of targeted
interventions that tackle both gender stereotypes and harmful
behaviors. It is essential to develop culturally tailored,
community-specific programs that challenge rooted beliefs and
cultivate a supportive environment for gender equality, particularly
among young people. A comprehensive challenging approach is vital,
encompassing the implementation of educational programs focused on
healthy relationships and consent, awareness campaigns through
social media, accessible support services, parental engagement in
discussions about VAW, collaboration with community organizations,
advocacy for gender equality and legal reforms, and the
establishment of safe spaces for open dialogue.

Research to inform policy and communities, while shedding light on
the underestimated aspects of the phenomenon, is also essential.
Future developments should focus on several key areas: we propose to
explore causal relationships in observational data using Directed
Acyclic Graphs (DAGs) in the context of Bayesian networks (BNs)
\citep{Castelletti2020} and implementing post-stratification
estimation \citep{Triveni2024} to enhance estimator efficiency and
provide more accurate insights into attitudes toward VAW. Finally, a
comparative analysis of the study's results with official
statistics, once available, will offer valuable context and
validation, enriching our understanding of the dynamics of the
phenomenon in the local communities.

\end{document}